\documentclass[twocolumn]{aastex631}

\newcommand{\DM}{\textit{Direct Method }}
\newcommand{\CM}{\textit{Calibrated Method }}
\newcommand{\BE}{\textit{Binding Energy }}

\usepackage{multirow}

\shorttitle{AASTeX v6.3.1 Sample article}
\shortauthors{Piacentino et al.}
\graphicspath{{./}{figures/}}
\usepackage{lineno}

\begin{document}

\title{Computational Estimation of the Binding Energies of PO$_x$ and HPO$_x$ (x=2,3) Species }

\author[0000-0001-6947-7411]{Elettra L. Piacentino}
\affiliation{Harvard-Smithsonian Center for Astrophysics, 60 Garden Street, Cambridge, MA 02138, USA }

\author[0000-0001-8798-1347]{Karin I. {\"O}berg}
\affiliation{Harvard-Smithsonian Center for Astrophysics, 60 Garden Street, Cambridge, MA 02138, USA }

\begin{abstract}

The distribution of molecules between the gas and solid phase during star and planet formation, determines the trajectory of gas and grain surface chemistry, as well as the delivery of elements to nascent planets. This distribution is primarily set by the binding energies of different molecules to water ice surfaces. 
We computationally estimated the binding energies of ten astrochemically relevant P-bearing species on water surface, we also validate our method for 20 species with known binding energies. We used DFT calculations (M06-2X/aug-cc-pVDZ) to calculate the energetics of molecules and water-molecule clusters (1-3 H$_2$O molecules) and from this determined the binding energy by comparing the complex and the separate molecule and cluster energies. We also explore whether these estimates can be improved by first calibrating our computational method using experimentally measured binding energies. Using the 20 reference molecules we find that the 2H$_2$O cluster size yields the best binding energy estimates and that the application of a calibration to the data may improve the results for some classes of molecules, including more refractory species. Based on these calculations we find that, small P-bearing molecules such as PH$_3$, PN, PO, HPO, PO$_2$ and POOH are relatively volatile and should desorb prior or concomitantly with water ice, while H$_2$PO, HPO$_2$, PO$_3$, PO$_2$OH can strongly bind to any hydroxylated surface, and will likely remain on the interstellar grains surface past the desorbtion of water ice. The depletion of P-carriers on grains constitute a pathway for the inclusion of Phosphorous molecules in planets and planetesimals. 

\end{abstract}
\keywords{Phosporous chemistry --- Binding Energy ASW --- Computational BE}

\section{Introduction}\label{Intro}
The relative abundance of chemical species in the ice and gas phase in astrochemical environments is of fundamental importance to predict the composition of planets and planetesimals \citep{marboeuf2014planetesimals,williams2011protoplanetary,2009A&A...505..183O}.
In the early stages of star formation the chemical distribution between ice and gas phase determines the chemical inventory that can be accessed for gas phase and surface chemistry on grains. In disks the balance between adsorption and desorption processes regulates which molecules can be incorporated into planets and planetesimals and which can only be delivered to planets through gas accretion. The temperature gradient present in disks results in the formation of condensation lines and therefore a radius dependent chemical distribution  \citep{bergin2015tracing,oberg2011effects,oberg2021astrochemistry}.
The location of these condensation lines depends on the strength of this physisorbtion interaction -  the binding energy (BE)- between a molecule and a solid surface (i.e. water ice, silicate, and carbonaceous grains) and consequently its availability in condensed or gaseous phase at a particular disk radius.

Among the biogenic elements, the distribution of phosporous (P) containing molecules between gas and solid phases, is perhaps most uncertain. The availability of P is key to the formation of several biotic molecules \citep{pasek2005aqueous} and it is quite abundant on Earth ( P/H $\sim$ 10$^{-3}$, \cite{fagerbakke1996content}). In comparison the cosmic abundance of P relative to hydrogen is much lower (P/H $\sim$ 2.57 x 10$^{-7}$, \citet{asplund2009chemical}).
Phosporous carried by PN and PO has been detected in the gas-phase around evolved stars \citep{tenenbaum2007identification,ziurys2007chemical,milam2008constraining} and in star forming regions (massive; \cite{fontani2016phosphorus,rivilla2016first}, low mass; \cite{bergner2019detection,yamaguchi2011detection,lefloch2016phosphorus}). In all these cases the P abundance is low accounting for a P/H of about 10$^{-10}$-10$^{-9}$\citep{rivilla2020alma,bergner2019detection,lefloch2016phosphorus,rivilla2016first}.
In circumstellar envelopes, such as around IRC +10216, phosphorous, carried by PH$_3$ and HCP, has been observed with abundances of $~$10$^-8$ with respect to molecular hydrogen \citep{agundez2012molecular}. This accounts for about 7\% of the phosphorous elemental abundance \citep{agundez2014confirmation}. These evidences suggest that the undetected phosphorous is likely incorporated into grains and that there must therefore be additional less volatile carriers of P in the interstellar medium.

The nature of less volatile P-containing compounds in the ISM is currently unclear, but Solar System studies may provide some clues.
Analysis of CI chondrites have shown an elemental P abundance similar to the solar phosphorus abundance \citep{lodders2003solar}. In stony meteorites most of the phosphorous is carried by Ca, Mg-phosphate minerals, while reduced phosphorous is more common in Fe-rich meteorites \citep{pasek2004quantitative}. Very recently volatile phosphorous, mainly carried by PO fragments, was detected on comet 67P/Churyumov–Gerasimenko during the Rosetta mission (\citep{gardner2020detection,rubin2019elemental,altwegg2016prebiotic}. This might suggest that also in the ISM, P is incorporated into a relatively refractory phase, which likely consists of species containing PO$_x$ moieties.

In this study we address the possible distributions of phosphorous oxides in the ISM and disks through a theoretical investigation of the binding energies of interstellar P carriers candidates with a PO$_x$ moiety. To determine relevant binding energies it is necessary to define a reasonable model system. Dust grains are composed of silicates aggregates \citep{jones2017global} or carbonaceous materials. In molecular clouds and other cold and dense interstellar and circumstellar environments, dust grains are coated with ices that formed by the freeze out of molecules in the gas phase which can chemically evolve via surface chemistry. Due to its high abundance, water is the major constituent of the icy surface \citep{boogert2015observations} in clouds and throughout star and planet formation. As such, water ice represents the most relevant surface for the evaluation of the binding energy of volatile and semi-volatile species. Additionally, the binding energies on water surface may also serve to estimate the Binding Energy (BE) on minerals as minerals surfaces are often hydroxylated \citep{landmesser1997interior,schaible2014hydrogen}.

The binding energy of stable species can be determined in the laboratory via Temperature Programmed Desorption (TPD) experiments \citep{chaabouni2018thermal,fayolle2016n2,collings2004laboratory}. In this study we use 20 astrochemically relevant species that have had their binding energies determined through TPD studies to evaluate our computational approach. 
These species are generally stable small molecules which include C, H, N and O atoms. Some example of experimentally measured binding energy of S-bearing species are also available (i.e. H$_2$S), while there is lack of information in regards to P-bearing molecules, especially unstable ones.

In the case of unstable and exotic species, the binding energy needs to be estimated computationally. 
A computational challenge is the accurate representation of the Amorphous Solid Water (ASW). Several studies have focused on a periodic representation of the solid water surface \citep{ferrero2020binding,zamirri2019quantum,karssemeijer2014diffusion,karssemeijer2014interactions}; In these studies the BE of a molecule is estimated from its interaction with a sizable ASW surfaces that contains multiple binding sites. As shown in \citet{ferrero2020binding} these approaches can yield a range of BE for each molecule dependent on the optimized binding site thus provides a distribution of binding energies to ASW for each molecule.

The ASW can also be approximated by medium to large sized water clusters. \citet{shimonishi2018adsorption} used 20 water molecules to define the the ASW geometry allowing for the definition of multiple binding sites.Very recently, \citet{2022ESC.....6.1286G} developed a computational method to build large (~1-200 H$_2$O) ASW cluster to determine accurate binding energy distribution ranges for molecules on ASW.

Other studies have focused on a small-cluster representation of the ASW which are more computationally affordable.
Although the use of small cluster does not provide for a comprehensive account of long range molecule-ASW interactions and for periodic variation of the ASW, it constitutes a lighter computational investment for each new molecule while still providing BE estimations well in range of both the periodic representation studies and the experimental value \citep{ferrero2020binding,wakelam2017binding,das2018approach}. 
The optimal size for small cluster is still debatable. \citet{das2018approach} tested cluster size of 1,3-6 H$_2$O and showed that the uncertainty on the BE is reduced as the cluster size increases.
On the other hand, single water cluster systems may provide similar uncertainties to BE estimated using periodic systems, when the calculated values are calibrated using experimental values \citep{wakelam2017binding}.

In this work we aim to computationally constrain the binding energies of 3 known ISM P-carriers as well as 7 proposed interstellar P-carriers over ASW. We will use two approaches described in detail in \S\ref{Methods}: direct calculation and calculations calibrated against experimental data on C, N, O and S-bearing molecules. The results of the two methods are presented in S\ref{Results} where we also discuss the reliability of these approaches. In section \S\ref{discussion} we present and comment on the application of these methods to P-bearing molecules. We include some astrophysical implications of our new binding energies for P-bearing molecules (\S\ref{Astro}). Finally, in \S\ref{conclusion} we summarize our findings.

\section{Methods}\label{Methods}

\subsection{Computational Details and Cluster size Choice}

When computing binding energies there are several aspect of the process, such as the computational tools, the molecular approximations, and data treatment, for which choices need to be made.

To model the binding interactions we chose to use electronic structure based methods. In particular, within the density functional theory (DFT) we chose to use the M06-2X functional for its good performances in modeling non-covalent interactions \citep{mardirossian2017thirty}. All calculations are run using Gaussian 16 suite of software \citep{g16} at the M06-2X/aug-cc-pVDZ level of theory \citep{zhao2008m06,kendall1992dunning,dunning1989gaussian} and included the optimization of the clusters geometry to a stationary point as well as a vibrational frequencies calculations for the identification of the energy minima.\footnote{Dataset is available at DOI:10.5281/zenodo.6551710} Additionally we tested the relative performances of Moller–Plesset (MP) methods \citep{frisch1990direct}, in particular at the MP2/aug-cc-pVDZ level of theory, for binding energy determination and we concluded that the two methods yields equivalent results ( Appendix \ref{A}). Previous work by \citet{wakelam2017binding} and \citet{ferrero2020binding} also chose the M06-2X functional for the estimation of binding energies.

Concerning the description of the surface-molecule interaction, we chose to focus our work on cluster systems rather then on a periodic representation of the water surface due to its computational affordability. The use of small cluster systems rather than a more extended periodic representation of the binding surface could lead to less accurate results.
In the work from \citet{ferrero2020binding} the authors show, however, that the binding energies calculated using small water clusters are comparable to more expensive and/or complex periodic calculations. 
In our study the cluster size was purposefully kept small in the attempt to minimize the computational cost to enable easy scaling to larger molecular data sets. Inspired by the work of \citet{das2018approach}, who found an improvement in accuracy going from the monomeric to the tetrameric representation of ASW while 5 and 6H$_2$O clusters did not further improve the accuracy, we evaluate the impact that the size of the water cluster has on the binding energies estimation by using 1H$_2$O, 2H$_2$O, and a 3H$_2$O water cluster sizes.
We also follow \citet{wakelam2017binding}, and explore whether directly calculated binding energies can be improved through calibration against experimental values using a molecular training set.
We extend this work by applying a calibration to the binding energies calculated using 1-3 H$_2$O water cluster sizes. 
To obtain a calibration set of molecules, we searched the literature for experimental data on binding energies to ASW surfaces and identified 20 molecules with well-defined experimental binding energies (Table \ref{Full-Table}).
The same literature data were also used to benchmark the performance of our computational results by direct comparison to the experimental values.

Concerning the uncertainties of the experimental binding energies, we apply a 10\% uncertainty to the BEs of species for which an uncertainty value of $<$ 10\% was reported in the literature and we conservatively consider an uncertainty of 30\% when the uncertainties were not reported in the literature (see Table \ref{Full-Table}).

The \DM, which estimates the binding energies from the energetics of 1H$_2$O, 2H$_2$O, and a 3H$_2$O water cluster sizes calculation is described in details in \S\ref{Method1:method}. \S\ref{Method2:method} describes the application of the \CM to the direct method data.

\begin{figure}[hb!]
  \centering
  \includegraphics[width=\columnwidth]{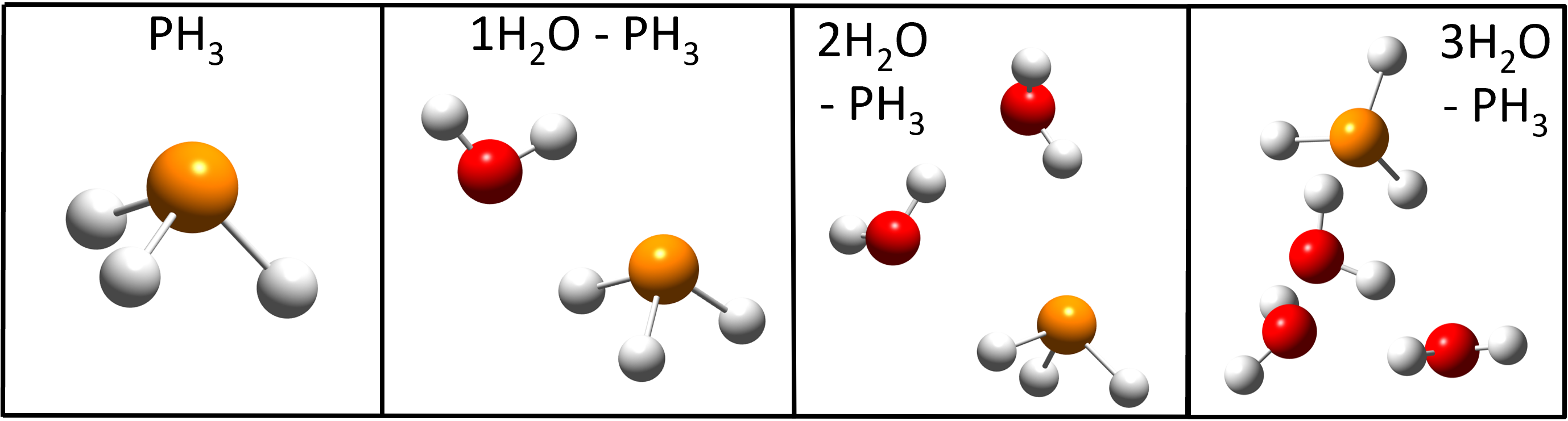}
 \caption{Example of optimized geometries of the species, 1H$_2$O systems,2H$_2$O systems, and 3H$_2$O systems used in this work.}
 \label{SAMPLE_SYSTEMS_PH3}
\end{figure}

\subsection{Direct Method}\label{Method1:method}
We first calculate the binding energies at different cluster sizes directly. The electronic energy of each molecule, water cluster, and water-molecule complex was calculated at the M06-2X/aug-cc-pVDZ level of theory for the 1, 2 and 3 H$_2$O cluster representation.
The geometry of the water-molecule complexes is optimized starting from a non-interaction configuration with the molecule placed at a minimum distance $>$ 5{\AA} from the water cluster. This is to ensure that no molecule-cluster interactions are present in the input geometry. For most molecules we performed a single optimization calculation (single initial configuration) which yielded geometries in agreement with the criteria described below. For molecules that allow multiple unique interaction configurations we repeated the optimization with 2-3 initial configurations. 

The following criteria were used to chose the representative geometry; In the 1H$_2$O clusters the representative geometry was chosen as the one where the main interaction was between H$_{H{_2}O}$ and the molecule. This is because the ASW surface is more rich in hydrogen than oxygen atoms and therefore the H$_{H{_2}O}$-molecule interaction is more likely to occur \citep{wakelam2017binding}. Generally the H$_{H{_2}O}$ interacts with an atom of the molecule but in the cases of unsaturated hydrocarbons the interaction is set between the H$_{H{_2}O}$ and the double or triple bond on the carbon chain. Exception to this criteria are CO$_2$, for which the main interaction is always set between the O$_{H{_2}O}$ and the C$_{CO{_2}}$, and CH$_3$OH which sees the main interaction occurring between the hydroxylic H$_{CH{_3}OH}$ and the O$_{H{_2}O}$. These exception in the final geometry are a results of the optimization calculation and are likely due to the nature of the molecules. The 2H$_2$O and 3H$_2$O complex geometries are chosen in a similar way while allowing for the second and third interaction to occur between the O$_{H{_2}O}$ and the molecules. An example of the optimized geometries are shown in Figure \ref{SAMPLE_SYSTEMS_PH3} for the molecule PH$_3$. A few more optimized geometries are shown in Appendix \ref{B}.

\begin{figure*}[!ht]
\centering
  \includegraphics[width=1\textwidth]{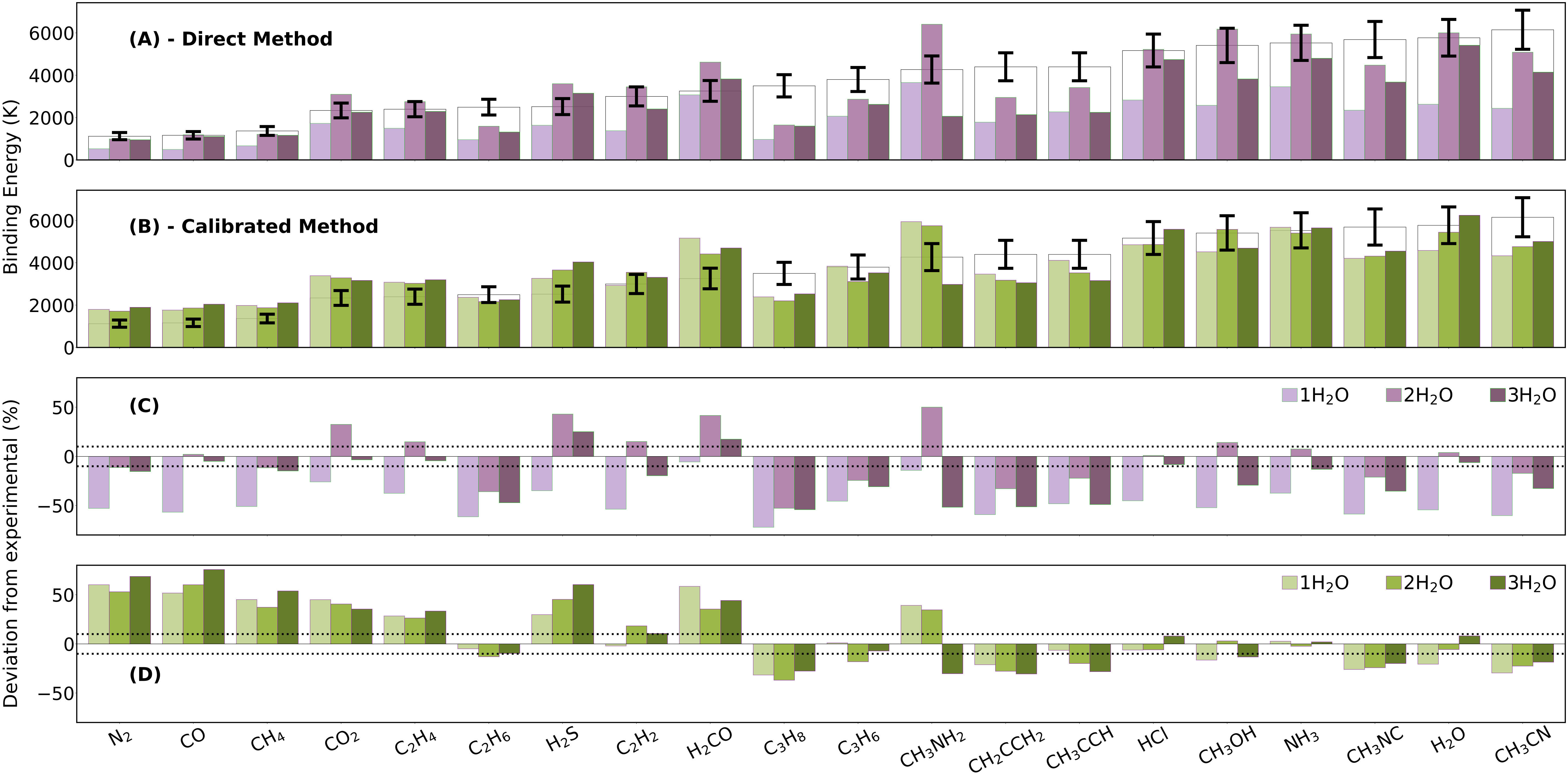}
 \caption{Binding energies (in K) of the 20 chemical species used in this study calculated using three different H$_2$O cluster sizes and  \DM (Panel A) and \CM (Panel B)in relation to the experimental BE of the same species (Black contour, 10\% error bars). The deviation from the experimental values are shown in panel C and D respectively.}
 \label{Deviation-plot}
\end{figure*}

In the \DM, the adsorption energies of the molecules in question is determined from the variation of the energy of an adsorbant molecule and a H$_2$O cluster (in our case constituting  1-3 H$_2$O water molecules) that arises when they are able to non-covalently coordinate with each other. The binding energy (BE) is calculated as follow:

\begin{center}
    \textit{BE}=E$_{complex}$-(E$_{molecule}$+E$_{H{_2}O cluster}$)\\ \label{Eq. 1}
\end{center}

Where E$_{complex}$ is the energy of one of the potential energy minima geometries of the water-molecule cluster when the species is physisorbed onto the surface of the cluster (2-5 {\AA}),  E$_{molecule}$ is the energy of the species alone, and E$_{H{_2}O cluster}$ is the calculated energy of the water cluster.

The calculated electronic energies are used directly without accounting for the zero point energy and the Basis Set Superposition Error (BSSE) similarly to \citet{wakelam2017binding}. \citet{wakelam2017binding} tested whether the inclusion of ZPE and or BSSE significantly affected the accuracy of the resulting fit and found that, in the dimer case, the inclusion of the corrections slightly reduced the goodness of the fit. This suggests that while the omission of either the ZPE or the BSSE may significantly affect the calculated energies the omission of both corrections in the estimation of binding energies yields values that better approximate the experimental BE \citep{das2018approach}. This errors compensation is likely just a fortuitous but nonetheless advantageous balance.
The \DM is evaluated for a reference set of 20 molecules (Table \ref{Full-Table}) and then applied to calculate the BEs of 10 P-bearing species (Table \ref{P-table}).

\subsection{Calibrated Method}\label{Method2:method}

\citet{wakelam2017binding} demonstrated that there is a systematic offset between calculated and experimentally determined BEs when the BEs are calculated using a single water molecule cluster. They also show that the accuracy of the calculated BE estimates can be improved if they are calibrated against experiments. \citet{wakelam2017binding} calculated, using the equation in section \ref{Eq. 1}, the  BE's of 16 molecules using a 1H$_2$O ASW representation to build a calibration curve against the experimental BE of each molecule. From the fit they evaluated the interaction correlation between a 1H$_2$O representation and the ASW.
Building upon the work of \citet{wakelam2017binding}, we extended the method by applying the \textit{Calibrated Method} to the 1-3 H$_2$O systems to evaluate the effects of the increased cluster size on the performances of the \CM. We also evaluate whether the use of the \CM  provides a significant improvement in the BE estimation over the \DM.

The calibration curves were built by fitting the 20 BE's (20 for each of the 3 water cluster systems) obtained using the \DM against the experimental BE's values assuming a linear relationship between calculated and experimental values. The fit was then applied to the BE's calculated with the \DM resulting in the \CM estimation of the BE values. 
The \CM was then applied to estimate the BEs of the P-species.

\section{Methods Validation}\label{Results}
Table \ref{Full-Table} summarizes the calculated \DM and estimated \CM binding energy values for the calibration molecules. Figure \ref{Deviation-plot} shows the binding energies calculated using the \DM and the \CM as well as the percentage deviation from the experimental values. Below we present these results in detail In \S\ref{discussion} we present the application of the two methods to our selection of phosphorous molecules.

 \begin{deluxetable*}{c c||ccc|ccc|cl}
\tablecaption{Calculated and experimentally determined binding energies (in K) for our reference molecules. \label{Full-Table}}
\tablehead{
\multicolumn{2}{c||}{\textit{M06-2X}}&\multicolumn{3}{c|}{\DM}&\multicolumn{3}{c|}{\CM}&\multicolumn{2}{c}{Exp.$^b$}\\
\multicolumn{2}{c||}{\textit{aug-cc-pVDZ}}	& 		&  &\multicolumn{1}{c|}{}& &&\multicolumn{1}{c|}{}&&\\
\multicolumn{2}{c||}{Species} & 1H$_2$O &2H$_2$O&\multicolumn{1}{c|} {3H$_2$O} & \colhead{1H$_2$O} &\colhead{2H$_2$O}  &\multicolumn{1}{c|} {3H$_2$O}&\colhead{}  &\colhead{}}
\startdata
1	&	N$_2$	&	530	&	998	&	953	&	1803	&	1723	&	1898	&	&	1125$^{[1,2]}$	\\
2	&	CO	&	504	&	1191	&	1107	&	1769	&	1867	&	2048	&	&	1165$^{[1,2,3,4]}$	\\
3	&	CH$_4$	&	672	&	1212	&	1168	&	1990	&	1882	&	2108	&	&	1370$^{[2,5]}$	\\
4	&	CO$_2$	&	1733	&	3104	&	2259	&	3395	&	3289	&	3170	&	&	2339$^{[4,5,6]}$	\\
5	&	C$_2$H$_4$	&	1498	&	2760	&	2294	&	3084	&	3034	&	3204	&	&	2400$^{[7]}$	\\
6	&	C$_2$H$_6$	&	960	&	1600	&	1321	&	2372	&	2171	&	2256	&	&	2495$^{[2,7]}$	\\
7	&	H$_2$S	&	1641	&	3606	&	3154	&	3273	&	3663	&	4042	&	&	2519$^{[8,9]}$	\\
8	&	C$_2$H$_2$	&	1388	&	3458	&	2413	&	2938	&	3552	&	3320	&	&	3000$^{[7]}$	\\
9	&	H$_2$CO	&	3076	&	4623	&	3832	&	5172	&	4419	&	4702	&	&	3260$^{[9,10]}$	\\
10	&	C$_3$H$_8$	&	976	&	1652	&	1606	&	2394	&	2210	&	2534	&	&	3500$^{[2,7]}$	\\
11	&	C$_3$H$_6$	&	2071	&	2871	&	2629	&	3842	&	3116	&	3531	&	&	3800$^{[7]}$	\\
12	&	CH$_3$NH$_2$	&	3660	&	6414	&	2065	&	5944	&	5750	&	2982	&	&	4269$^{[11]}$\\
13	&	CH$_2$CCH$_2$	&	1792	&	2957	&	2145	&	3472	&	3180	&	3059	&	&	4400$^{[7]}$\\
14	&	CH$_3$CCH	&	2280	&	3422	&	2251	&	4118	&	3525	&	3162	&	&	4400$^{[7]}$\\
15	&	HCl	&	2836	&	5225	&	4743	&	4854	&	4866	&	5589	&	&	5170$^{[12]}$	\\
16	&	CH$_3$OH	&	2584	&	6183	&	3826	&	4520	&	5579	&	4696	&	&	5410$^{[13]}$\\
17	&	NH$_3$	&	3464	&	5949	&	4805	&	5685	&	5405	&	5649	&	&	5530$^{[14]}$	\\
18	&	CH$_3$NC	&	2352	&	4484	&	3680	&	4213	&	4315	&	4555	&	&	5686$^{[15]}$\\
19	&	H$_2$O	&	2633	&	6007	&	5416	&	4586	&	5448	&	6245	&	&	5773$^{[16]}$	\\
20	&	CH$_3$CN	&	2445	&	5093	&	4148	&	4337	&	4769	&	5010	&	\omit&	6150$^{[15]}$\\
\enddata
\tablecomments{References are 1. \citet{fayolle2016n2}; 2. \citet{smith2016desorption}; 
3. \citet{collings2003laboratory}; 4. \citet{noble2012thermal}; 5. \citet{he2016binding}; 6. \citet{galvez2007study}; 7. \citet{behmard2019desorption}; 8. \citet{wakelam2017binding}; 9. \citet{penteado2017sensitivity}; 10. \citet{noble2012desorption}; 11. \citet{chaabouni2018thermal}; 12. \citet{olanrewaju2011probing}; 13. \citet{bahr2008interaction}; 14. \citet{hama2013surface}; 15. \citet{bertin2017nitrile}; 16. \citet{fraser2001thermal}; 
17. \citet{fraser2001thermal}}
\end{deluxetable*}

\begin{figure}[hb!]
  \centering
  \includegraphics[width=\columnwidth]{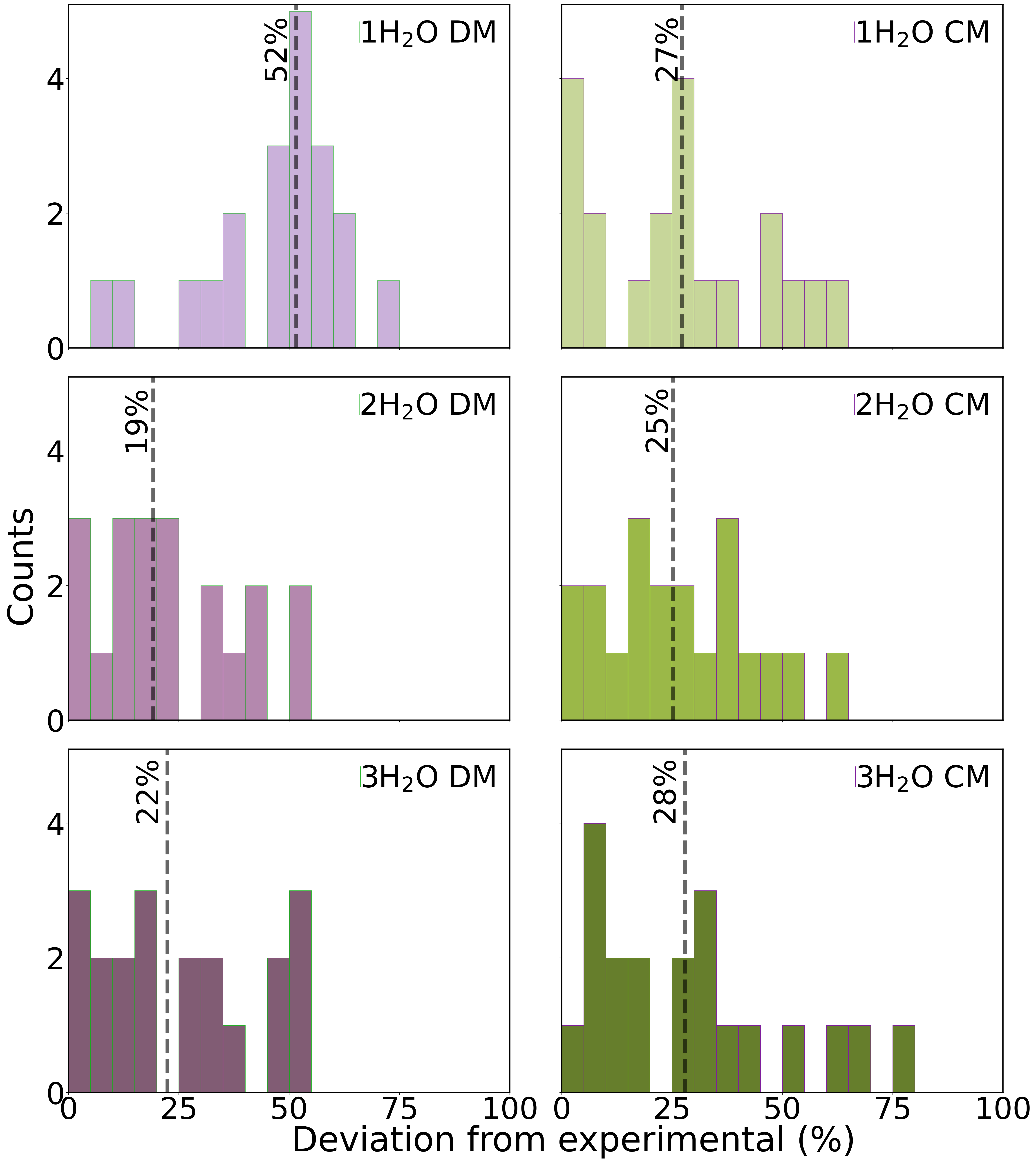}
 \caption{Bin distribution of the deviation from the experimental value of the computational binding energies of 20 reference molecules calculated using the \DM (DM) and the \CM (CM)  using a 1-3 H$_2$O representation of the ASW surface. The dotted lines highlight the median values of the distributions.}
 \label{HIST}
\end{figure}

\subsection{\DM H\texorpdfstring{$_2$O}. Cluster vs. ASW Binding Energies} \label{Method1:resuls}

Figure \ref{Deviation-plot} and Table \ref{Full-Table} show that the accuracy of the calculated \DM binding energies is highly variable both between different molecules, and when using different cluster sizes. Calculations accounting for 1H$_2$O interaction performed the worst, with deviations as high as 72\% and a median deviation of 52\%, while the 2 and 3-H$_2$O clusters performed better with deviations between a few and 54\% and median deviations of 19 and 23\% respectively. This deviations can be compared to typical experimental errors of 10\%. The performance improvement when increasing the cluster size from one to two or three H$_2$O molecules is visualized in Fig. \ref{HIST}, which show histogram plots of the deviations in percentage from experimental values. In other words, there is a real increase in performance when increasing the cluster size from one to two, but not when increasing it to three for our sample of molecules. This suggests that increasing the cluster size beyond two H$_2$O molecules may be of limited value when the focus is to computationally determine the mean value of a molecule binding energy using the cluster approach. However, this needs to be confirmed for a larger and more diverse sample of molecules, as well as for a larger range of cluster sizes.
While the small cluster approach is an useful and computationally inexpensive tool to determine the mean \BE values, it provides limited information on the \BE distribution and therefore cannot be applied in cases when the whole range of the \BE distribution is of interest.

Interestingly the performance of the \DM approach appears to depend on the strength of the binding interaction. Figure \ref{Deviation-plot} shows that for molecules up a volatility of $\sim$ 3000K the median deviations are only $\sim$15\% for the 2-H$_2$O and 3-H$_2$O cluster calculations, while there is almost a 1.5 factor increase when considering the less volatile species. 
There are also evidences that the accuracy of the \DM approach depends on the chemical nature of the molecule. The binding energies prediction for hydrocarbons (i.e. C$_2$H$_6$, C$_3$H$_8$, C$_3$H$_6$, CH$_2$CCH$_2$, CH$_3$CCH) seem to be less accurate than for other molecules with a deviation from the experimental values between 25 and 53\% for the 2H$_2$O cluster size (Fig. \ref{Deviation-plot}, panels A and C).

Finally we note that there are a handful of molecules for which the deviation from experimental values increases between the 1H$_2$O and 2H$_2$O clusters, namely:
CO$_2$, H$_2$S, H$_2$CO and CH$_3$NH$_2$. These molecules warrant further investigation, since this may be revealing something interesting about their interactions with water ice. For now we simply note that in each of these cases the binding energy proceeds from being slightly underpredicted in the case of 1H$_2$O, to overpredicted for the 2H$_2$O cluster.

\begin{figure}[hb!]
  \centering
  \includegraphics[width=\columnwidth]{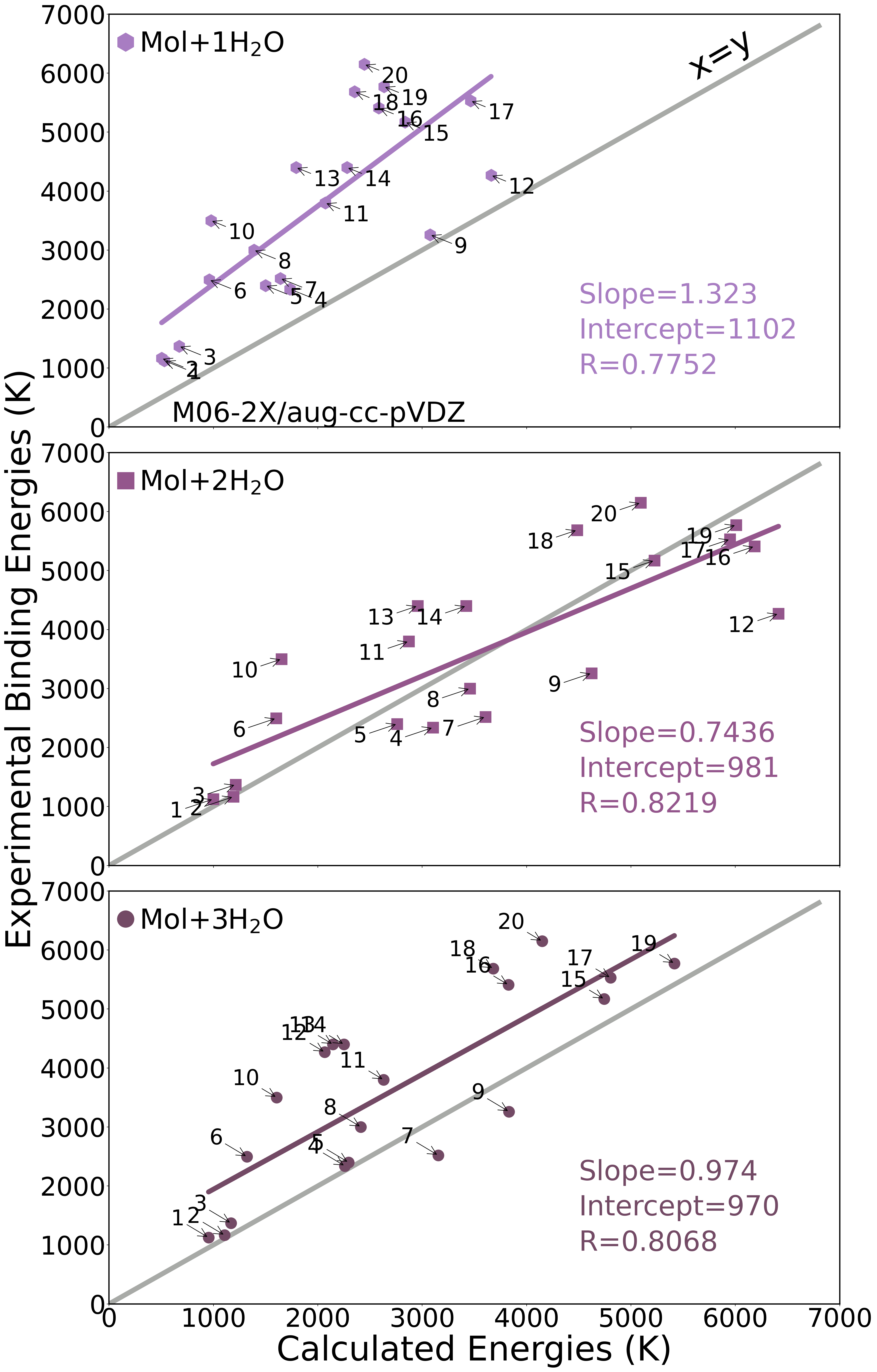}
 \caption{Linear fit of the 20 binding energies calculated using the \DM for each of the 1-3H$_2$O cluster systems against the experimental values from the literature. The data point numbers correspond to the indexes of the molecules in Table \ref{Full-Table}}
 \label{CAL_CU_F}
\end{figure}

\begin{figure}[hb!]
  \centering
  \includegraphics[width=\columnwidth]{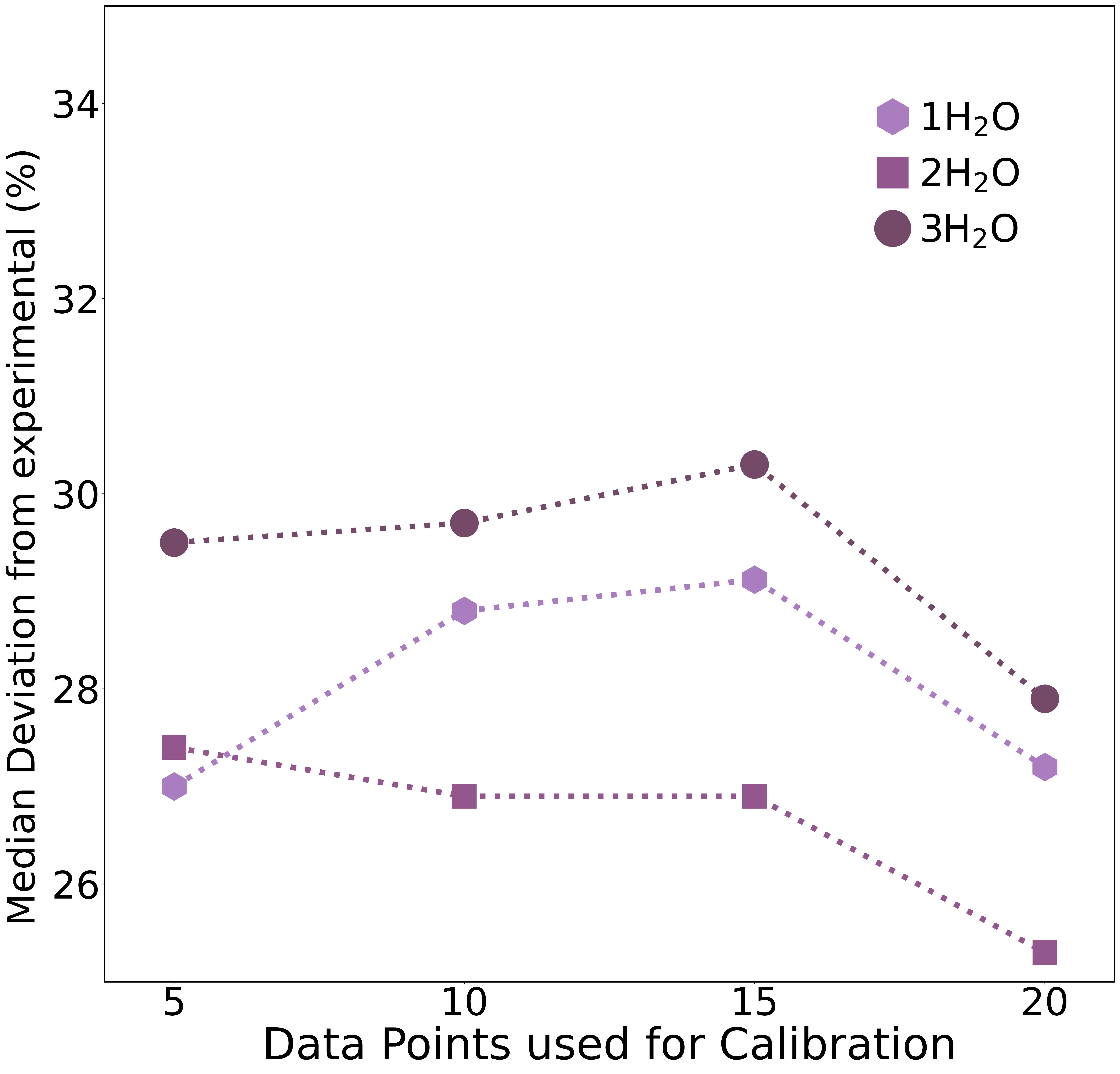}
 \caption{Median of the deviation from the experimental values for the 1-3 H$_2$O systems calculated after applying a 5,10, 15, and 20 data points calibration.}
 \label{CAL_TREND}
\end{figure}

\subsection{Consideration on the Direct Binding Energies Calculation}\label{Method1:Discussion}
A first measure of the accuracy of the \DM binding energy calculations is the deviation from experimental values. 
As reported above the deviation from experimental binding energies decreases from 52 to 19\% (Figure \ref{HIST}) when increasing the cluster size from 1 to 2H$_2$O, and no further improvement is seen when increasing the cluster size from 2 to 3H$_2$O (22\% median deviation). The improvement when going from 1 to 2/3H$_2$O clusters is in line with the results from \cite{das2018approach}, who found an improvement in the deviation from the experimental values from $~$40 to $~$25\% when increasing the cluster size from 1 to 3H$_2$O molecules (2H$_2$O cluster calculations were not included in the study).
\cite{das2018approach} also reported a consistent underestimation of the BE values compared to the experimental values, that is present for both the monomeric and the trimeric representation of the ASW, additionally showing a trend of progressively less negative estimations as the cluster size increases. Our calculations for the 1 and 3H$_2$O clusters show a similar trend where the 3H$_2$O system generally underestimate the experimental binding energies while providing a substantial improvement form the binding energies derived from the 1H$_2$O system.

However, we find that the 2H$_2$O system does not fit in the same trend, underestimating only 45\% of the binding energies studied. The randomness of this distribution along with the good absolute median uncertainty of 19\% suggest that, among the cluster size and geometries that are considered in this study, the 2H$_2$O description of the ASW surface provides the best binding energies prediction.

The improvement in the results when using 2/3H$_2$O clusters vs. 1H$_2$O is intuitively due to the accounting for additional interactions between the molecule and the water cluster. The presence of more water molecules allows for additional binding constrains between the molecule and the water cluster yielding a binding geometry that better resemble the binding configuration On the ice.
We can test this intuition by interrogating the molecule-cluster systems in detail. We find that the addition of a second water molecule results in a geometry where the primary water molecule can bond more strongly with the molecule in question. Since in an ice system there are always neighboring molecules, accounting for this distortion is important to produce accurate binding energies.
Not accounting for this geometrical distortion yields underestimated BE: this is consistent with the underestimation (Fig. \ref{CAL_CU_F}) of the binding energies in the 1H$_2$O system where not all the fundamental interaction between the species and the ASW can be taken into account. 

\begin{figure}[hb!]
  \centering
  \includegraphics[width=0.5\columnwidth]{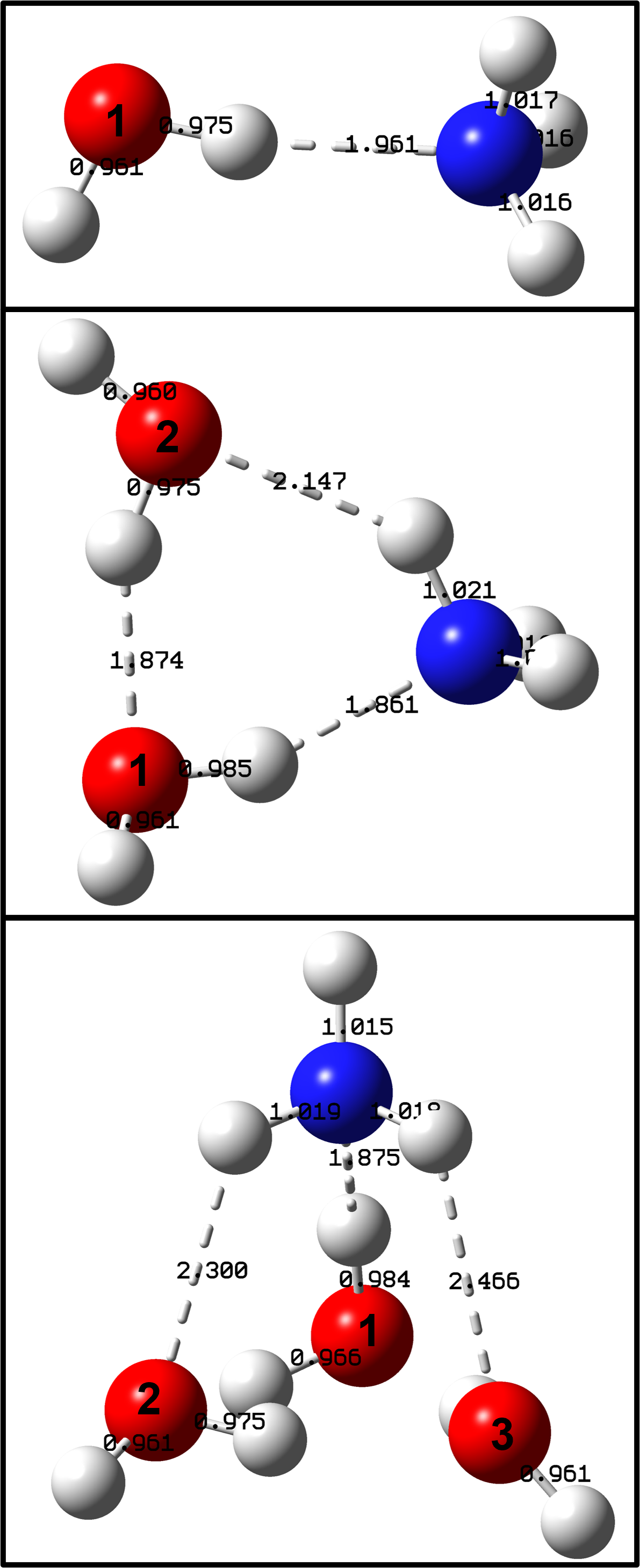}
 \caption{Optimized binding geometries (M06-2X/aug-cc-pVDZ) of NH$_3$ in combination of the 1, 2, and 3 H$_2$O representation of the ASW surface. Bond lengths and long range interactions (dotted lines) are in {\AA}.}
 \label{NH3}
\end{figure}

Figure \ref{NH3} shows an example (NH$_3$) of the effects on the molecules binding environment caused by the second H$_2$O. 
The main interaction between the ASW and NH$_3$ is between the nitrogen on NH$_3$ and one of the water hydrogen. In the case of 2H$_2$O the interaction distance is shorter (1.96 {\AA} and 1.86 {\AA} respectively in the 1 and 2 H$_2$O systems). A consequent elongation of the O-H bond in the primary water molecule is also observed.
The second water molecule does not interact as strongly with the NH$_3$ but provides an additional anchoring point (2.15 {\AA}) resulting in the reduction of the N--(H-O)$_{H{_2}O}$ angle from 171$^{\circ}$ to 159$^{\circ}$. A similar behavior is generally observed across the studied molecules.

Following the same intuition as above, we should observe an improved BE accuracy when increasing the H$_2$O cluster size from 2 to 3H$_2$O, but this is not what we find.
The presence of the third water molecule introduces additional structural constraints (2.46 {\AA}, Figure \ref{NH3}) which cause a weakening of the primary and the secondary interaction between the molecule and the water. 
This results in a binding energy prediction for the 3H$_2$O system very similar to the 2H$_2$O system with the absolute error for ammonia going from 8\% in the 2H$_2$O to 13\% in the 3H$_2$O.
The observed lack of increased precision when increasing the cluster size from 2 to 3 H$_2$O molecules is surprising. It is typically expected for larger clusters to increase the accuracy of the binding energy estimation as more long range interactions can be taken into account and less unique geometries become available.
One possible explanation is that the use of a greater number of water molecule introduces significant freedom in regards to the arrangement of the water molecules themselves and this may produce geometries at odds with ASW. 
In our study we found that the binding energy estimations are dependent on the functional group that the molecule use to bind to the water cluster.This is most apparent when comparing the isomers CH$_3$CN and CH$_3$NC.
For CH$_3$CN and CH$_3$NC the binding energy estimates reflect the interactions of the water with either the N or C atoms that are terminal to the molecule. In the case of CH$_3$NC we have that the interaction distance is 2.18{\AA} which becomes 2.07{\AA} for CH$_3$CN consequently increasing the binding energy of the molecule. This is consistent with the experimental binding energy and it reflects the affinity of the water for the functionalities present in the molecules. The binding distance is further shortened for molecules having a terminal oxygen, in the case of H$_2$CO for example, the O$_{CH{_2}O}$- H$_{H{_2}O}$ distance is 1.98\AA.

We find a similar functionality dependency of the binding energy to water in cases where the main binding interaction occurs between the H$_{H{_2}O}$ and the double/ triple bond in the molecules.
In the case of CH$_2$CCH$_2$, and CH$_3$CCH, for example, we find that CH$_3$CCH binds strongly to the water cluster when compared to CH$_2$CCH$_2$. 
This is because the triple bond in CH$_3$CCH constitute a better binding functionality for hydrogen than the double bond in CH$_2$CCH$_2$. 
This effect of the molecule saturation on the binding energy prediction is observed also in the C2 and C3 series of hydrocarbons with smaller binding energies as the saturation of the molecule increases (Table \ref{Full-Table}). This same trend has also been studied experimentally by \citet{behmard2019desorption} who observed a similar trend showing that the binding energy of C2 and C3 hydrocarbons decreases with the saturation of the molecules (i.e. BE$_{C{_2}H{_2}}$ $>$ BE$_{C{_2}H{_4}}$ $>$ BE$_{C{_2}H{_6}}$)
In general, the application of the \DM to hydrocarbons seem to yield a worse approximation of their \textit{Binding Energies} compared to other molecules in the study. The median deviation for non-hydrocarbons is 16\% and for hydrocarbons 23\%, in the 2H$_2$O \DM calculation.
The underestimation of the BE of hydrocarbons may be associated to the poor natural affinity that this class of molecules have toward water.
In such cases the size of the cluster may be more relevant than for other classes of molecules and the use of a periodic representation of the ASW may be of aid. The presence of a matrix may help producing a tighter packing of the water around the molecule, which would result in a binding structure that better resemble the experiments. The poor constrain that the \DM provide for hydrocarbons suggests that this class of molecules may also especially benefit from the calibrated method. 

\subsection{\CM Improvements Over the \DM} \label{Method2:resuls}
We next evaluate whether the results obtained through the direct method can be improved upon application of the \CM using the same 20 molecules. Figure \ref{CAL_CU_F} shows the linear fit of the \BE calculated using the \DM and the experimental values. The calculated binding energies using 1 and 3H$_2$O generally underestimate the binding energy. This effect is more pronounced and less uniform for the 1H$_2$O compared to the 3H$_2$O data set. The use of 2H$_2$O appears to instead produce a close to random scatter around the expected values. We note, however, that as we go from more to less volatile molecules, the 2H$_2$O cluster method seems to be systematically over-predict the binding energies.

We find that the \CM binding energy predictions are very similar to each other for all the cluster sizes. By contrast to the \DM, we find no improvement in the median deviation from the experimental value as the cluster size is increased (Fig. \ref{HIST}). For 1, 2 and 3H$_2$O clusters we find median deviations of 27\%, 25\% and 28\%, and deviation ranges of 1-60\%, 2-60\% and 2-69\%, respectively. This implies that the \CM generally achieves a higher level of accuracy when considering single water molecule clusters, but a comparable level of accuracy when considering larger clusters in comparison with the \DM.

Using the \CM we see an opposite dependence of the results accuracy with the molecule volatility compared to the \DM (Fig. \ref{Deviation-plot}, panel D). For volatile molecules, having \BE $<$3000K, the median deviation from the experimental values calculated using the \CM-2H$_2$O cluster size is 37\%. The median deviation is reduced to 19\% for the less volatile group of molecules, which implies that the calibrated method outperforms the direct method for the more refractory molecules. It also appears to do better with the 2 and 3C hydrocarbons compared to the direct method.

Additionally we tested the importance of the number of calibration points used to estimate the correlation for uncertainties minimization; We performed 1000 random selections of 5, 10 and 15 molecules from our list of 20 molecules and for each set we used the resulting fits to determine the binding energies of the molecules not included in the selections. We finally calculated the median of the errors across the 1000 draws for each of the 5, 10 and 15 molecules sets (Fig. \ref{CAL_TREND}). The median deviation decreases as the calibration set increases, but the improvement is small: about 2\%. Though it will be interesting to revisit the \CM approach with more experimentally determined values, we may already be close to the limit where the peculiarities of each molecule-H$_2$O system dominates the calculated uncertainty. 
\subsection{Utility of Calibrating Calculated Binding Energies?}\label{Method2:Discussion}
The application of the \CM is more or less advantageous in dependence of the cluster size used and the volatility of the molecules in exam.
Without a calibration the 1H$_2$O BEs are too inaccurate to be useful, and we hence recommend that such a calibration is always used for 1H$_2$O BE calculations. The application of the \CM to the 2 and 3H$_2$O cluster sizes do not appear to contribute to the accuracy of the results except for some specific classes of molecules.

The improvements to the predicted BEs observed when applying the calibration to the less volatile species in contrast with the loss of accuracy for highly volatile molecules ($<$3000K) suggests that the \CM  does not, on average, improve the results. However, we suspect that the calibration could improve the results if applied in a more targeted way. More experimental binding energy values are needed to separately calibrate volatile and non-volatile species.

In a similar way, we noticed that for some class of molecules such as hydrocarbons the binding energies seem to be difficult to estimate compared to other types of molecules. 
This reduced accuracy for hydrocarbons, further suggests that a blind calibration using heterogenous collections of molecules might not be the best choice for all class of compounds. 
It is possible, however, that the use of more narrowly defined families of molecules, also in combination with the use of bigger cluster sizes, could improve the results.

The availability of more experimentally measured binding energies will be crucial for the effective computational prediction of the binding energies of species that show chemical class dependencies and the consequent application of computational methods to estimated the binding energies of species that are not of easy access in a laboratory.
\begin{figure*}[th!]
  \centering
  \includegraphics[width=\textwidth]{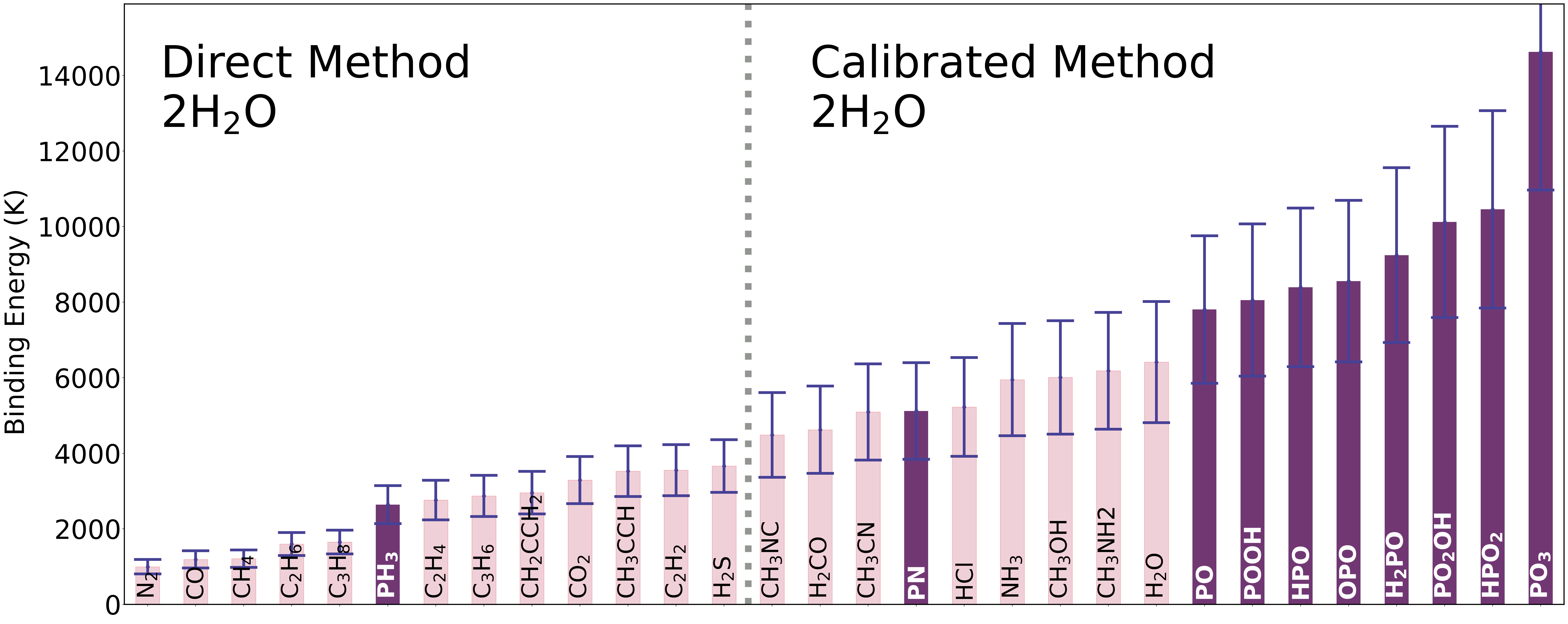}
 \caption{Binding Energies values, in K, calculated using the \DM (low range \BE)  and \CM (high range \BE)  in combination with the 2H$_2$O cluster of both the reference species and the P-bearing species. We reported uncertainties of 19\% for the \DM resutls and of 25\% for the \CM results ( Figure \ref{HIST}). Species are plot in order of increasing calculated BE.}
 \label{Figure-P}
\end{figure*}

\subsection{Result Summary and Recommendation }
Overall we find that the \DM results obtained using the 2H$_2$O cluster size provide binding energies with the lowest median uncertainty on the whole range of reference molecules of 19\%. The application of the \CM only improved on the 2H$_2$O cluster size for the least volatile molecules and for 2C and 3C hydrocarbons.
The cut-off appears to be around a BE of 3000K, but a larger molecular reference set is needed to explore this further.

Given these results, we recommend using the \DM 2H$_2$O for volatile species and the \CM 2H$_2$O for semi-volatile and more refractory species to determine the mean \BE value using small water clusters.

 \begin{deluxetable*}{c||ccc|ccc}
\tablecolumns{7}
\tablecaption{Binding energies of selected P-bearing in K \label{P-table}}
\tablehead{
\textit{M06-2X}	&	\multicolumn{3}{c|}{\multirow{2}{*}{\DM (K)}}	&	\multicolumn{3}{c}{\multirow{2}{*}{\CM (K)}}\\
\textit{aug-cc-pVDZ}	&		& &		&		&		&\\
\hline								
Species	&	1H$_2$O ($\pm$ 51.6\%)	&	2H$_2$O ($\pm$ 19.2\%)	& 	3H$_2$O	($\pm$ 22.4\%) &	1H$_2$O	($\pm$ 27.2\%)&	2H$_2$O ($\pm$ 25.3\%)	&	 3H$_2$O ($\pm$ 27.9\%)}
\startdata
PH$_3$	&	1117	&	2642	&	1228	&	2579	&	2945	&	2166	\\
PN	&	2168	&	5119	&	2376	&	3970	&	4787	&	3284	\\
PO	&	3791	&	9176	&	9221	&	6117	&	7804	&	 9951	\\
HPO	&	4176	&	9964	&	5304	&	6626	&	8390	&	6136	\\
OPO	&	4327	&	10186	&	11031	&	6826	&	8555	&	11714	\\
POOH	&	625	&	9512	&	4551	&	9382	&	8054	&	5403	\\
PO$_2$OH	&	7254	&	12290	&	25502*	&	10698	&	10119	&	25809*	\\
HPO$_2$	&	7457	&	12740	&	15388	&	10967	&	10454	&	15958	\\
H$_2$PO	&	9118	&	11112	&	8858	&	13164	&	9244	&	9597	\\
PO$_3$	&	11039	&	18343	&	30441*	&	15706	&	14621	&	 30619*	\\
\enddata
\tablecomments{* These structures react with the ASW to form a PO$_4$ moiety. The energy reported refer to a chemisorption events.}
\end{deluxetable*}

\section{Application of the Methods to P-molecules}\label{discussion}

\subsection{BE of P-Bearing Molecules}\label{P-results}
We calculated the binding energies of 10 phosphorous molecules (Table \ref{P-table}) using the \DM and the \CM. The uncertainties on the calculated binding energies values are derived from the median deviation from the experimental value obtained respectively for each of the methods (Fig. \ref{HIST}). When estimating the uncertainties for the \CM of the P molecular BEs below, we use the median deviations for the fiducial 20-molecule calibration set.
As expected from the calibration set the \DM 3H$_2$O systems estimations provides, for most of the P-bearing molecules, binding energies values that are in between the values obtained using the 1H$_2$O and the 2H$_2$O \DM calculations (Table \ref{P-table}). Exceptions are OPO, and HPO$_2$, for which the \DM 3H$_2$O binding energy values are higher than the 2H$_2$O prediction. Additionally, the geometry optimization of PO$_2$OH and PO$_3$ with 3H$_2$O results in the coordination of the P-species to one oxygen from the water cluster yielding a PO$_4$ moiety (See Appendix \ref{C}). This complexation prevents the calculation of the physisorption energy for these two P-molecules for the 3H$_2$O cluster system. We note that the complexation is observed only for the 3H$_2$O cluster setting, in all other cluster size neither covalent interactions nor deformation of the water cluster geometries are observed.

Similarly to what observed for the reference set of molecules we find that the also in the case of P-bearing molecules the application of the \CM estimates binding energies values that are equivalent across the three cluster systems. We also find similarities between the fiducial set and the P-bearing set of molecules when comparing the two methods performance for each cluster size.
In the case of 1H$_2$O cluster size we find that, similarly to the fiducial set of molecules, the \CM estimations have higher BE values than the \DM by a factor of $\sim$ 2.4 for low desorbing species, and $\geq$ 1.5 for the less volatile species. 
The 2H$_2$O cluster size yields a smaller discrepancy between the \DM and \CM prediction compared to the 1H$_2$O cluster size with the 2H$_2$O estimation yielding values within 15\% of the \DM estimation. 
In the cases when the 3H$_2$O clusters size did not yield additional coordination chemistry, the \CM estimation relates to the \DM estimation by a 1.8 and 1.2 factor respectively for volatile and less-volatile species.

In figure \ref{Figure-P} the binding energies values calculated for the P-bearing species are shown in relation to the binding energies calculated for the reference set of molecules using the 2H$_2$O system.
With the exception of PH$_3$, PO, and PN the estimated binding energies of the P-bearing species are found to be quite high (BE $\sim$8000K). With a most of the molecules exceeding the range of the fiducial set of molecules (PO$_2$OH, HPO$_2$, H$_2$PO, and PO$_3$) with BE$>$9000K.

Considering the results obtained for the calibration set of molecules, we divided the P-bearing species in three groups based on their binding energies. The first group consist of PH$_3$ alone, which is the only highly volatile molecule with a binding energy below 3000K when calculating energies for the P molecules. PH$_3$ binding energy is estimated using the \textit{Direct Method}. PN, PO constitute the second group, defined by binding energies similar to, or lower than water ice \citep{fraser2001thermal}. For these molecules we recommend using the \CM results. In the highest range of binding energies we find the refractory group constituted by HPO, OPO, POOH, H$_2$PO, PO$_2$OH, HPO$_2$ and PO$_3$ for which the binding energies exceed 8000K. For these too we recommend using the \CM 2H$_2$O but caution that the results are more uncertain since they extend beyond the calibration set.

\subsection{P-bearing Binding Energies Calculation}\label{P-Discussion}
We discuss the P molecule results by volatility grouping. The \textit{volatile} species, PH$_3$, PN, and PO, binding energies fall within the range of the experimentally determined binding energies that we evaluated our methods against. We therefore expect that our error estimations for these species are reliable.
The PH$_3$ binding energy has been previously computed by \citet{nguyen2021experimental} at 1813–2690K, this range is consistent with our calculation of 2642 ${\pm}$ 19\%. Our values are also in agreement with the binding energy range calculated, using a 20 water molecules cluster, by \citet{molpeceres2021computational} who reports 2189K as the average binding energy and with the computational BE reported by \citet{viana2015interaction} of 3000K calculated using a 2 water molecules cluster at the CCSD(T) and MP2 level of theory. Experimental works have shown that phospine's thermal sublimation occurs at around 60K \citet{turner2015photoionization}, consistent with a binding energy of $\sim$1800K using the formalism of \citet{hollenbach2008water}. In summary, our computational method appears to be accurate for phosphine. 

To our knowledge, there have not been experimental or computational studies on PO, PN.
Next we turn to the P-species with calibrated 2H$_2$O cluster BEs above 8000K, which makes them effectively refractory in astrophysical environments. The calculated BEs for these species present two complications; First of all they fall outside of our calibration range and their BE error bars are therefore more uncertain. Second, in interstellar regions, they are not expected to desorb off water ice.
However, we argue that the sublimation temperature on silicate grain should not substantially deviate from those expected on ASW. It has been shown that the binding energy of volatile species on ASW and on silicate surface falls in the same energy range \citep{suhasaria2017thermal}. In the case of CO$_2$ and CO the experimentally measured BE on silicates differs from the BE on water by less than 20\% \citep{noble2012thermal} 
This is likely due to the reactivity toward hydrogen of silicate surfaces, which results in them hosting -OH functionalities. This results in a binding behavior similar to the one of water \citep{landmesser1997interior,schaible2014hydrogen}.

We also found evidence that P-bearing species containing three oxygens can be further coordinated by a water molecule to form the PO$_4$ moiety when enough water molecules are present in the environment. Furthermore, the high BE found for OPO and HPO$_2$ using the 3H$_2$O cluster size hints to possibility that these molecules might coordinate with the water molecules in the cluster to form PO$_3$ species. The study of these species' reactivity may especially benefit from the use of bigger cluster size and from the use of dynamic models, this to better define the role that the ASW has in catalyzing the formation of additional PO bonds. We have not further explored this aspect in this work but, if this reactivity is proven to be viable in astrochemical environment it could provide a plausible pathway for the formation of complex and even more refractory P-bearing species. 

\subsection{Astrochemical Implications} \label{Astro}

The abundance of phosphorous compounds detected in the gas phase at various stages of cloud evolution varies significantly.While only 1\% of the expected phosphorous has been detected in star forming regions \citep{bergner2019detection,rivilla2016first,rivilla2020alma}, phosphorous, in its ionic form, has been detected with solar abundances in diffuse clouds in the ISM \citep{1978ApJ...219..861J,2006ASPC..348..480L}. 
This abundance discrepancy indicates that, during star and planet formation the majority of the phosphorous is depleted on icy grains in semi-refractory molecular carriers that have not yet been well constrained.

The majority of the P-bearing molecules that we explored in this work are found to be more refractory than water. This suggests that we should expect a semi-volatile to refractory phosphorous reservoir that remains in the solid-phase well after water sublimation. In disks, such species would remain solid interior to the water snowline. 

The abundance of the PO-bearing species simulated in this work will largely depend on their condensed phase formation chemistry and on the specific environmental conditions. In the solar system, organic phosponic acids\citep{cooper1992alkyl} as well as Ca-phosphate \citep{2014GeCoA.131..368L} have been detected on the Murchison meteorite suggesting the possibility for the existence of a rich phosphorous chemistry in condensed phase. We are currently investigating different scenarios computationally to determine the fraction of locked P that is attainable at different ISM conditions (Piacentino et al. in prep.).

In comets, (H)$_x$PO$_y$ species have been previously considered as possible phosphorous carriers \citep{rivilla2020alma}. Although the investigation on the ROSINA data collected on comet 67P did not provide the direct detection of (H)$_x$PO$_y$ species, it showed the presence of PO fragments which were attributed solely to PO molecules \citep{rivilla2020alma}.
As the fragmentation pattern of (H)$_x$PO$_y$ species is not well constrained, it is also possible for the detected PO signal to include a contribution due to the fragmentation of bigger phosphorus molecules. 
In either case we agree with \citet{rivilla2020alma}'s speculation that the PO signal is due to P-bearing molecules that were locked in the grains early on during star formation.

In our calculations we find that PO$_2$OH and PO$_3$ can chemisorb on the water surface to coordinate with an additional oxygen atom. This can indicate the tendency of these PO-bearing species to further react with water molecules to form the likely more refractory phosphate moiety. We speculate that the easiness with which phosphates may form could, depending on the environmental condition, lead to even a larger fraction of the phosphorous to be locked on grains in phosphate form.

In conclusion, while (H)$_x$PO$_y$ molecules have not yet being directly detected, the clues that we have indicate that these high desorbing species are good candidates for phosphorous carriers during star and planet formation, and may be the starting point of phosphate formation through their interaction with water ice.

\section{Summary and Conclusions}\label{conclusion}
We explored the performances of a direct ab-initio H$_2$O cluster calculation, and a calibrated version of the same, for binding energy estimations. We tested our methods using 20 molecules for which the experimental binding energies are well constrained in the literature and then applied these methods to 10 P-bearing molecules. We found that:
\begin{enumerate}
    \item The \DM-2H$_2$O cluster-method combination performs better than any other method/cluster size that we evaluated. It provides a quick binding energy estimation that does not seem to carry systematical errors.
    \item The application of the \CM improves the binding energy estimation for less volatile molecules while reduces the estimation accuracy for highly volatile molecules. This suggest that a targeted selection of the calibration set may be needed.  
    \item While the application of the \CM improves the results, the estimation of the binding energy for hydrocarbons seems to be difficult. This suggests that the use of an heterogeneous calibration set may not be optimal for all class of molecules. A functional group-based study may help highlight the effect of molecular proprieties on the BE estimation.
    \item The application of these computational methods to astrochemically interesting PO-containing species show that most of these species are more refractory than water.
    \item The presence of semi-refractory PO-containing species in disk beyond the water snowline could help explain the depletion of P in the ISM and it would supply a pathway for the inclusion of phosphorous in planets and planetesimals.
\end{enumerate}

This work was supported by a grant from the Simons Foundation 686302, KÖ. and by an award from the Simons Foundation 321183FY19, KÖ.

\bibliography{bib.bib}{}
\bibliographystyle{aasjournal}

\begin{appendix} \label{appendix}
\section{Selection of Model Chemistry} \label{A}
\begin{figure}[h]
  \centering
  \includegraphics[width=\textwidth]{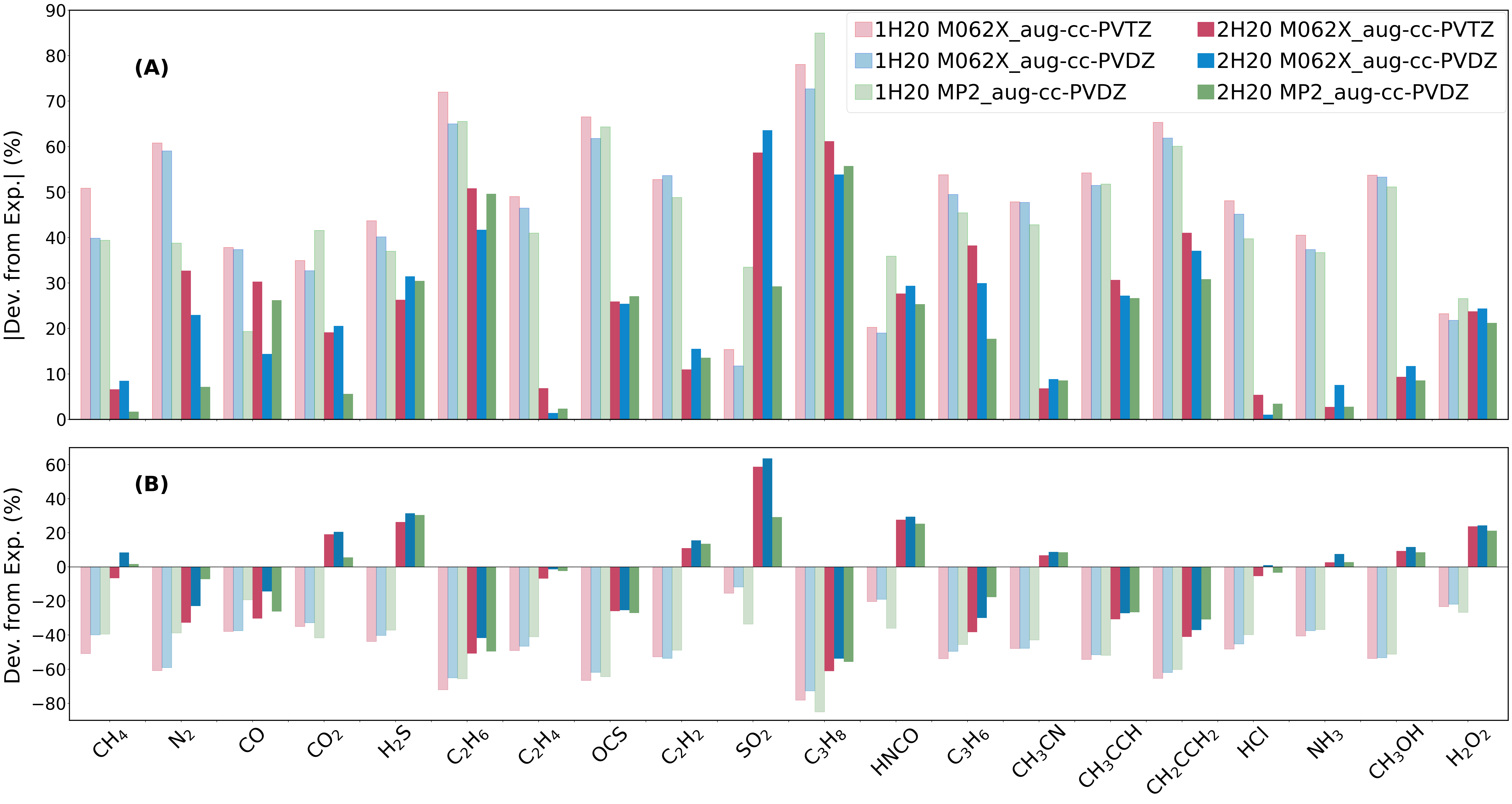}
 \caption{Comparison of the deviation from the experimental binding energy predicted by 3 computational models, M06-2X/aug-cc-pVTZ,M06-2X/aug-cc-pVDZ, and MP2/aug-cc-pVDZ, in combination with the 1H$_2$O and 2H$_2$O representation of the ASW surface. Panel (A) shows the absolute value of the deviation from the experimental values. Panel(B) reports the raw calculated values}
 \label{modelscomp}
\end{figure}

We also tested the model chemistry impact on the methods performances comparing the M06-2X results to similarly obtained BE values using an \textit{ab initio}, namely MP2 \citep{frisch1990direct}, model in combination with the same double zeta basis set. We also compare the M06-2X/aug-cc-pVDZ and M06-2X/aug-cc-pVTZ \citep{kendall1992dunning} performances in evaluating the binding energy.
As only a minimal variation in the BE calculated using the \DM at the M06-2X/aug-cc-pVTZ, M06-2X/aug-cc-pVDZ, and MP2/aug-cc-pVDZ is observed(Figure\ref{modelscomp}), we chose, for clarity, to limit out study to solely the M06-2X/aug-cc-pVDZ model chemistry.
We found that the binding energy estimations depend only slightly on the model chemistry used, but the variation is larger when the cluster size is increased from 1 to 2H$_2$O regardless of the model used. In a few cases - N$_2$,CO, HNCO, and SO$_2$ - there are real differences for different model chemistries within the same cluster size, which we speculate  are due to molecular peculiarity that we have not further investigated.  In either case, even for these molecules the differences are within the reported uncertainties ($\leq$ 20\%) justifying the use of a single model chemistry in the main section of the paper 

\section{Binding geometries of small hydrocarbons} \label{B}
\begin{figure}[h]
  \centering
  \includegraphics[width=\textwidth]{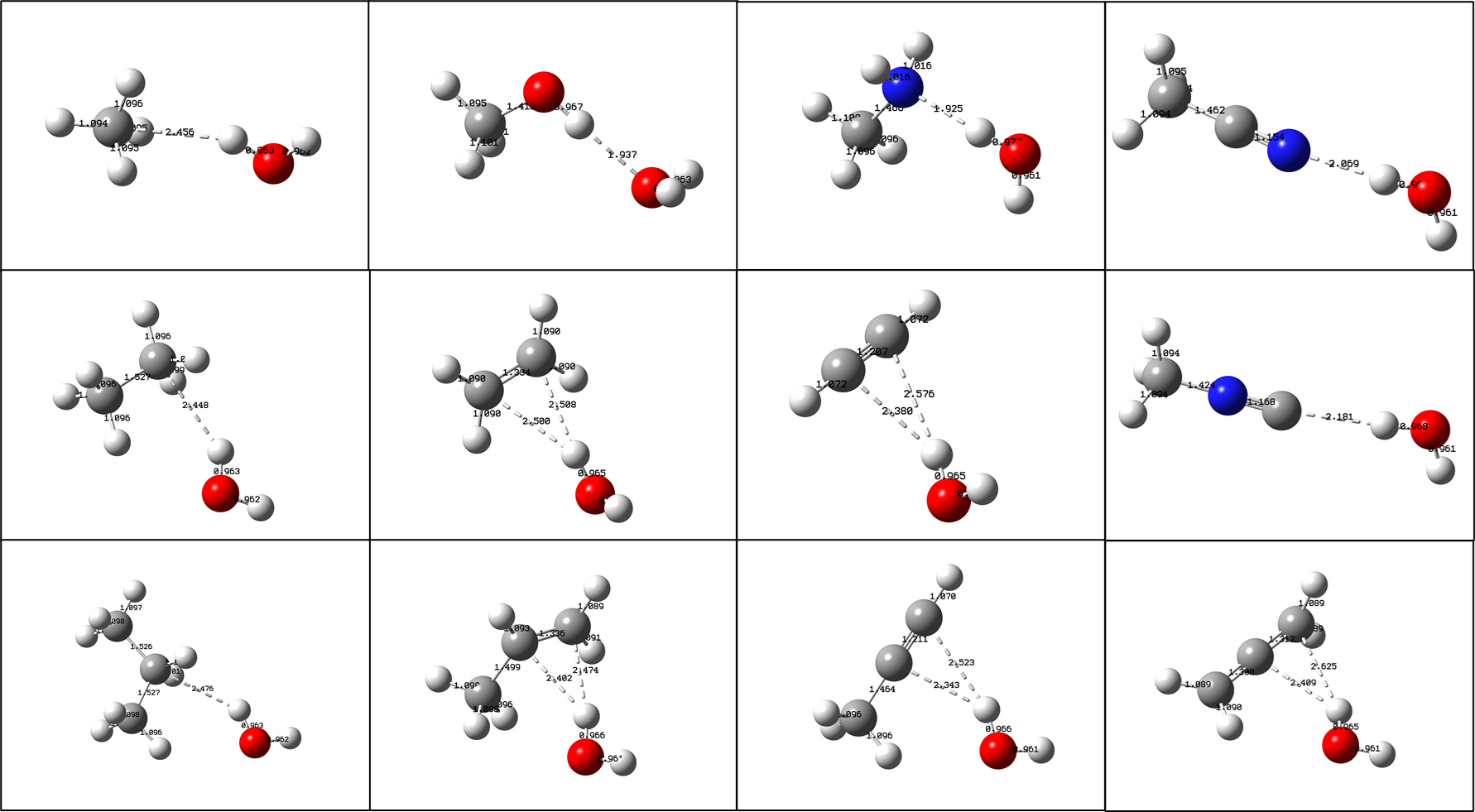}
 \caption{Optimized binding geometries (M06-2X/aug-cc-pVDZ) of small hydrocarbons with the 1 H$_2$O representation of the ASW surface. Bond lengths are in {\AA}.}
 \label{CO}
\end{figure}

\newpage
\section{PO\texorpdfstring{$_3$}. Optimization}
\label{C}
\begin{figure}[h]
  \centering
  \includegraphics[width=0.7\textwidth]{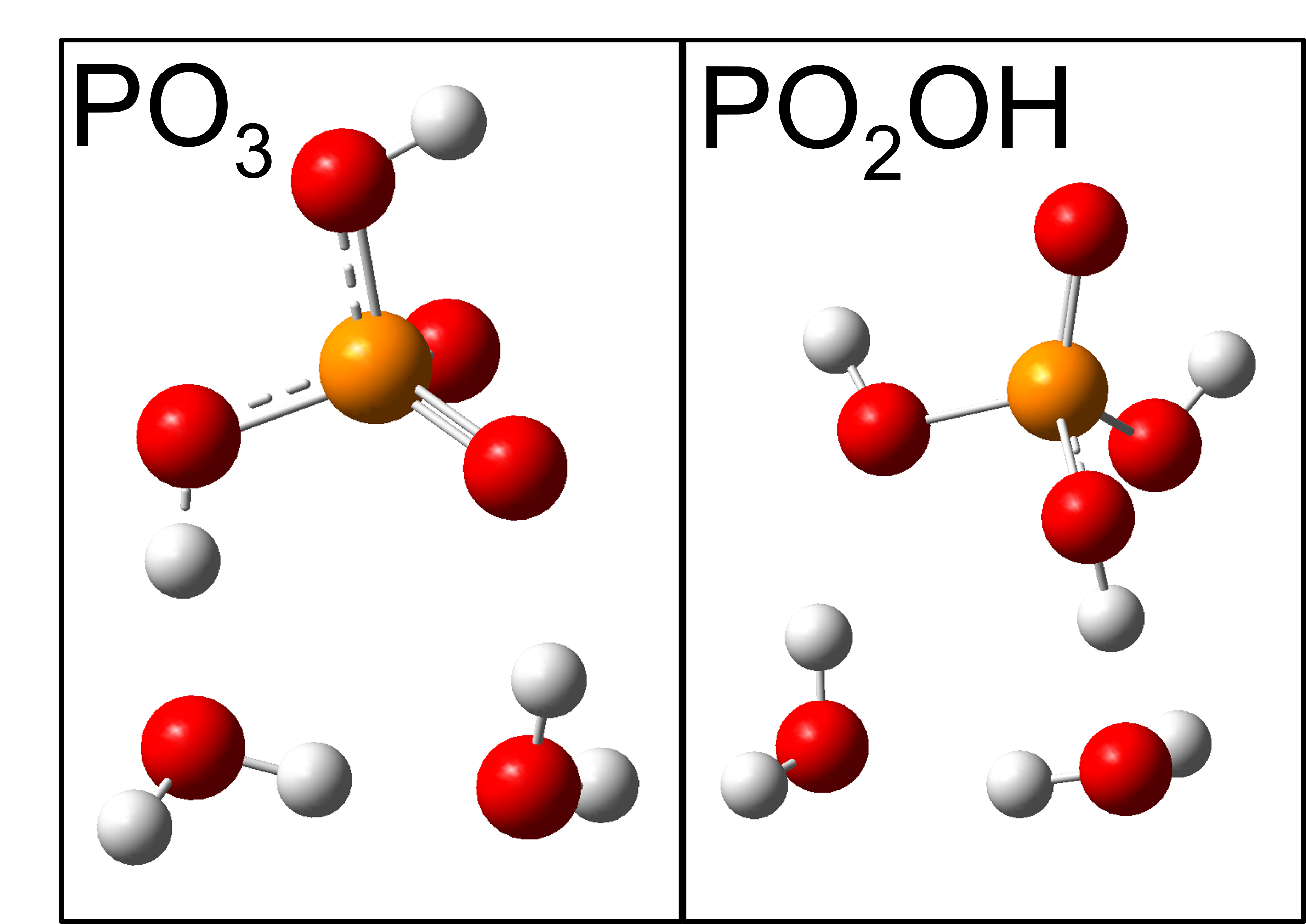}
 \caption{Optimized binding geometries (M06-2X/aug-cc-pVDZ) of PO$_3$, and PO$_2$OH with the 3H$_2$O cluster. This local minima shows formation of additional coordination between the P-species and one of the water of the cluster. This does not exclude the existence of physisorbtion configurations and it does not necessarily imply reactivity. Additional investigation which are beyond the scope of this work are needed to better understand this behavior.}
 \label{PO3}
\end{figure}

\end{appendix}

\end{document}